\documentclass[preprint, prX]{revtex4}

\usepackage{amssymb}
\usepackage{graphicx}
\begin{document}
\title{Epitaxial graphene morphologies probed by weak (anti)-localization}
\author{A. Mahmood}
\author{C. Naud}
\email{cecile.naud@grenoble.cnrs.fr}
\author{C. Bouvier}
\author{F. Hiebel}
\author{P. Mallet}
\author{J.-Y. Veuillen}
\affiliation{ Institut N\'eel, CNRS and Universit\'e Joseph Fourier,
BP 166, 38042 Grenoble Cedex 9, France}
\author{D. Chaussende}
\author{T. Ouisse}
\affiliation{Laboratoire des Mat\'eriaux et du G\'enie Physique,
CNRS UMR5628 - Grenoble INP, Minatec 3 parvis Louis N\'eel, BP257,
38016 Grenoble, France}
\author{L. P. L\'evy}
\email{laurent.levy@grenoble.cnrs.fr}
\affiliation{ Institut N\'eel,
CNRS and Universit\'e Joseph Fourier, BP 166, 38042 Grenoble Cedex
9, France}

\begin{abstract}
We show how the weak field magneto-conductance can be used as a tool
to characterize epitaxial graphene samples grown from the C or the
Si face of Silicon Carbide, with mobilities ranging from 120 to
12000 cm$^2$/(V.s). Depending on the growth conditions, we observe anti-localization and/or localization which can be
 understood in term of weak-localization related to quantum interferences.
The inferred characteristic diffusion lengths are in agreement with the scanning tunneling
microscopy and the theoretical model which describe the ``pure'' mono-layer and bilayer of graphene
 [MacCann {\it et al},. Phys. Rev. Lett. {\bf 97}, 146805 (2006)].
\end{abstract}


\maketitle

\vspace{2mm}

\section{Introduction}
Considerable progress has been achieved in the synthesis of
two-dimensional graphene. Since the seminal
works \cite{Novoselov05,Zhang05} which used exfoliated graphite
flakes transferred onto SiO$_2$ substrates, full wafers of epitaxial
graphene can now be grown by high temperature graphitization of
Silicon Carbide (SiC) crystals starting either from their Carbon or
Silicon face \cite{Berger04}. More recently, MBE growth on SiC
substrate have been achieved \cite{Moreau10} and CVD synthesis of large area graphene
films have also been carried out on the surface of transition metals in
high vacuum \cite{Coraux08} or at ambient pressure \cite{Kim09,Li09}.
The subsequent transfer of such CVD graphene films to a large variety of substrates is now controlled.
These synthesis methods are scalable and offer some real
perspectives for micro-electronic applications. A number of
characterization techniques are available for the grown layers: STM,
AFM, Raman, TEM/SEM and photo-emission have proven their usefulness.
On the other hand, the relationship between the growth conditions,
the film morphologies and the electronic properties have not yet
been systematically investigated \cite{Robinson09, Low12, Tanabe10,
 tersoff11, Yu10, Lee11, Jobst10, Creeth11, lara11}. This correlation is 
 important since epitaxial graphene presents characteristic defaults which 
differentiates it from the  ``pure''  free graphene.

\begin{figure}[!t]
\includegraphics[bb= 0 0 797 582,width=0.7\columnwidth]{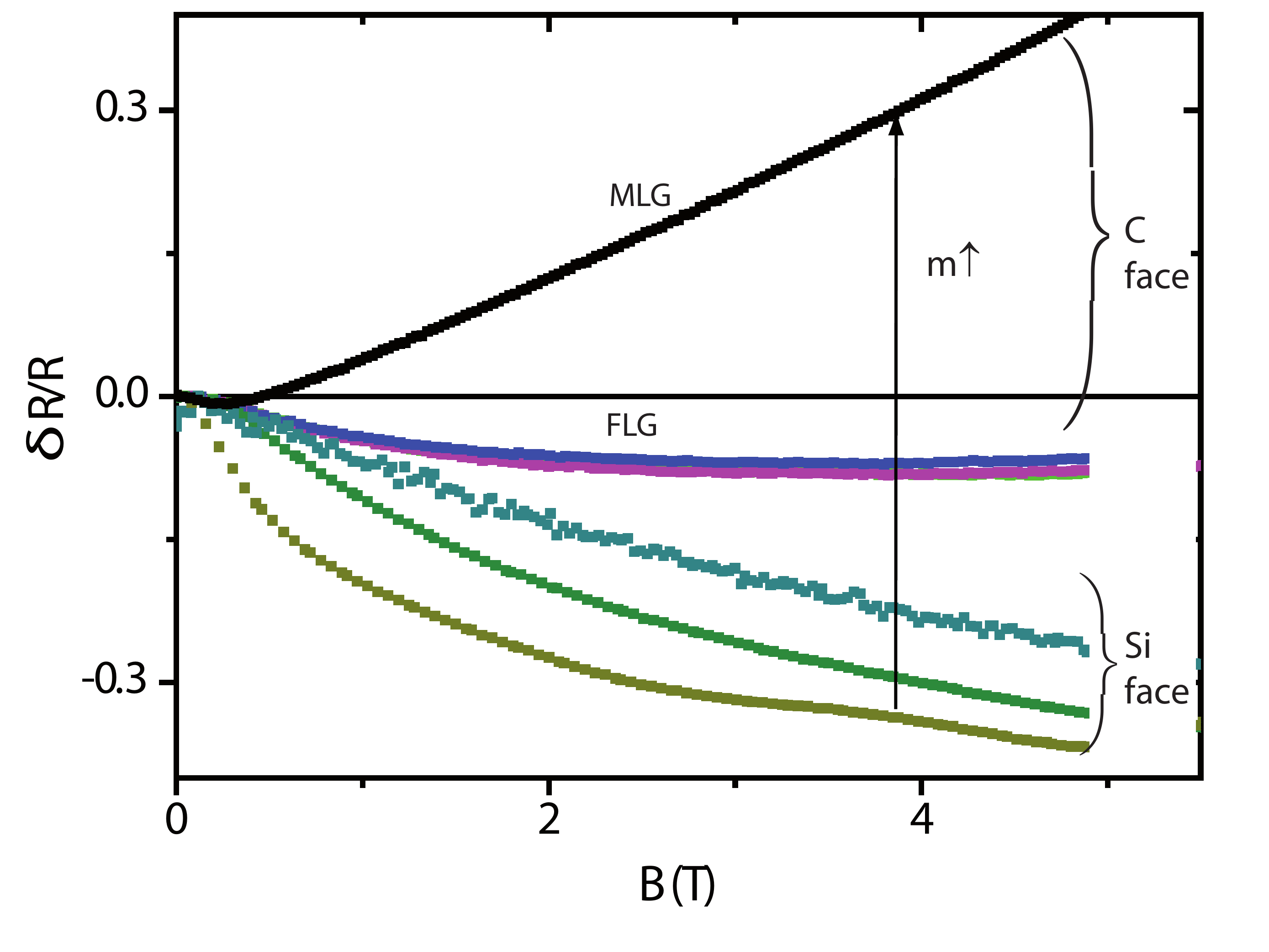}
\caption{Overview of the magnetoresistance traces on a large field scale (5T)
of 6 graphene epitaxial samples grown in different conditions and measured at 4K.  A positive
magnetoresistance is observed for the highest mobility (m) samples
(grown from the SiC C-face), with different slopes for few layers (FLG)
(number of layers $< 10$) and multilayer (MLG) (number of layers $>> 10$) graphene, while negative
magnetoresistance are observed on samples grown from the SiC
Si-face. All these traces are analyzed quantitatively in this paper and related
to the surface morphology.}
\label{Magneto-resistance-overview}
\end{figure}

In this paper, low field magneto-resistance is used to correlate the
transport properties, the
growth conditions and the morphologies of epitaxially-grown graphene
films elaborated from the different surfaces of 6H-SiC.  The films
studied have been grown with different graphene layer numbers, both
from the Si and C terminated faces, some in ultra-high vacuum other
in inert atmospheres. Depending on the SiC polytype and on the
growth conditions, distinct surface morphologies can be observed
which lead to very different magnetoresistance behaviors (see
figure \ref{Magneto-resistance-overview}).  Exploiting the unique
features of interference phenomena present in magneto-transport,
electronic properties can be related to the surface morphologies.

\section{Localization and anti-localization in graphene}
Low field magneto-resistance is a sensitive probe for electronic
transport as it measures the effect of quantum interferences along
closed paths \cite{Bergmann82, Bergmann84, kmellnitskii80}. Depending on the closed loop size, the
interferences can be constructive or destructive. For very small
loops, it has been demonstrated both
theoretically \cite{McCann06,Kechedzhi07} and
experimentally \cite{Wu07,Tikhonenko09} that interferences between
identical time reversed paths are destructive in graphene leading to
a negative magnetoconductance (positive magnetoresistance), which is
characteristic of anti-localization of electron waves. For graphene,
electron wavefunctions have four components and may be characterized
by two additional quantum numbers: the isospin and the pseudospin.
The isospin measures the relative wavefunction amplitude on the
equivalent sites (A-B) in graphene unit cell, while the pseudospin
measures to which band valley $K+$ or $K-$ the quantum
states belong \cite{Castro-Neto09}. Antilocalization is a
characteristic feature of graphene as the isospin (collinear to
momentum) undergoes a full rotation on a closed loop, changing the
wavefunction sign and so forbidding the backscattering. As the loop size increases, scattering mechanisms
lead to additional rotations of the isospin, as well as to the
scattering between different valley states, such that the pseudospin
need not be preserved on long paths. Two lengthscales characterize
this diffusion: the pseudospin is controlled by the intervalley
diffusion length $L_i$, and $L_*$ the intravalley diffusion length
controls the isospin random diffusion. The ``overall'' effect of
these processes on the interferences along time-reversed paths is to
change their sign back to the ``normal'' positive
magneto-conductance due to coherent backscattering \cite{McCann06,Kechedzhi07} observed in other
two-dimensional systems. Eventually, for extremely long paths (of
length greater then $L_\varphi$, the phase coherence length) and/or
high temperatures, inelastic scattering kills interferences. The
beauty of quantum interference is that a characteristic magnetic
field can be associated to every loop size, when half a flux quantum
is threaded within the loop area: hence the magnetic fields
$B_{\varphi,i,*}=\frac{\Phi_0}{4\pi L_{\varphi,i,*}^2}$ can be
associated to the lengthscales $L_{\varphi,i,*}=\sqrt{D\tau_{\varphi,i,*}}$ respectively.

\begin{table}[!b]
\caption{Characteristics and extracted parameters of the epitaxial
samples studied.}
\begin{tabular}{|l|c|c|c|c|c|c|c|} \hline \hline
Face & growth & layer \# & $\mu$(m$^2$/(V$\cdot$s)& D(m$^2$/s) &
$L_\varphi$(4K) & $L_i$ & $L_*$ \\
 \hline C & UHV & $\le 5$ & 0.018 &
0.0025 & 72 nm & 40 nm & 26 nm
\\ 
\hline
C & Ar & $\approx 50$ & 1.2 & 0.159 & 740 nm & &  \\
\hline Si & UHV & 2 & 0.012 & 0.0022 & 140 nm & 30 nm & 18 nm \\
\hline \hline
\end{tabular}
\label{Tab-samples}
\end{table}

It is useful to recall some of the general features of epitaxial
graphene on SiC. Graphene layers can be grown
by Si sublimation at high temperature \cite{Hass08a}. Electrical conduction is
known to be dominated by the completed layers closest to the
interface \cite{Berger06}. When growing from the SiC C-face, there is
a rotation between successive layers which effectively decouples the
layers \cite{Hass08b,Faugeras08}.  This is to be contrasted from
graphene layers grown from the Si face, where Bernal stacking breaks
the symmetry between the two (A-B) carbon sites \cite{Hass08a}. In
both cases, completed layers are continuous and ripples cover the
SiC vicinal steps. Depending on the growth condition, folds are also
observed. On ripples or folds, there is a local stretching of the
graphene bonds. The other types of known defects arise at the
graphene/SiC interface. Defects far from the graphene layer (at distance $d \gg
a$, where $a$ is graphene lattice constant) do not break the A-B
symmetry and contribute only to intravalley elastic scattering.
Sharp potential variations ($d \approx a$), may break the
graphene A-B symmetry locally. This scattering potential is
time-reversal even and affects simultaneously the isospin and
pseudospin and contribute both to the inter and intravalley
scattering \cite{McCann06,Kechedzhi07}. Ripples and folds stretch
bonds and contribute equally to inter and intra valley scattering.
Finally, trigonal warping contributes only to intravalley
scattering \cite{McCann06,Kechedzhi07}.

These scattering processes govern the crossover from localization at
low field to anti-localization at high field. The quantum
correction \cite{McCann06,Kechedzhi07} to the magneto-conductance
$\delta G^{mono}= \delta G_i+\delta G_*$ of a single graphene layer can be
split between the intervalley
\begin{equation}
\delta G_i = \frac{e^2}{\pi \hbar} \left[
F\left(\frac{B}{B_\varphi}\right)- F\left(\frac{B}{B_\varphi +2
B_i}\right)\right] \label{intervalley}
\end{equation}
and the intravalley contributions
\begin{equation}
\delta G_* = -\frac{2e^2}{\pi \hbar} \left[
F\left(\frac{B}{B_\varphi+B_i+B_*}\right)\right],
\label{intravalley}
\end{equation}
where the function $F(x)=\ln(x)+\Psi(1/2+1/x)$, and $\Psi$ is the
digamma function.

When the A-B symmetry is fully broken (for a bilayer on Si face,
the two sublattices A and B are no more equivalent) the
intravalley contributions have the opposite sign $\delta G^{bi}= \delta
G_i-\delta G_*$: the magnetoconductance increases monotonously with
field and antilocalization disappears \cite{McCann06,Kechedzhi07}.

\section{Results and discussion}

Figure \ref{Magneto-resistance-overview} gives an overview of all the
magnetoresistance behaviors observed at 4K over a broader range of
magnetic field. The transport properties have been measured using 
contacts pads evaporated through a mask with a Van der Paw geometry (surface of $500 \mu m^2$). 
Four points DC measurements were performed down to $4.2 K$.
  Films grown from the SiC C-face have larger
mobilities and positive magnetoresistance at high fields.  Thick
films grown from the C-face (top trace) show a linear
magnetoresistance at larger fields. The positive magnetoresistance observed at high field
for SiC C-face samples can be contrasted with the SiC Si-face
graphene which have lower mobilities and a negative
magnetoresistance at all fields.

These differences can be understood in terms of weak localization
and anti-localization in graphene. Using the appropriate weak-localization formulae, the intervalley,
intravalley and phase coherence length can be obtained by fitting
the magnetoconductance curves. All the results are summarized
in table \ref{Tab-samples} and are discussed in the rest of this article.

\subsection{\label{sec:Si-face}Magneto-conductance on the Si-face of 6H-SiC}

\begin{figure}[!pth]
\includegraphics[bb = 0 0 1734 2176,width=0.48\columnwidth]{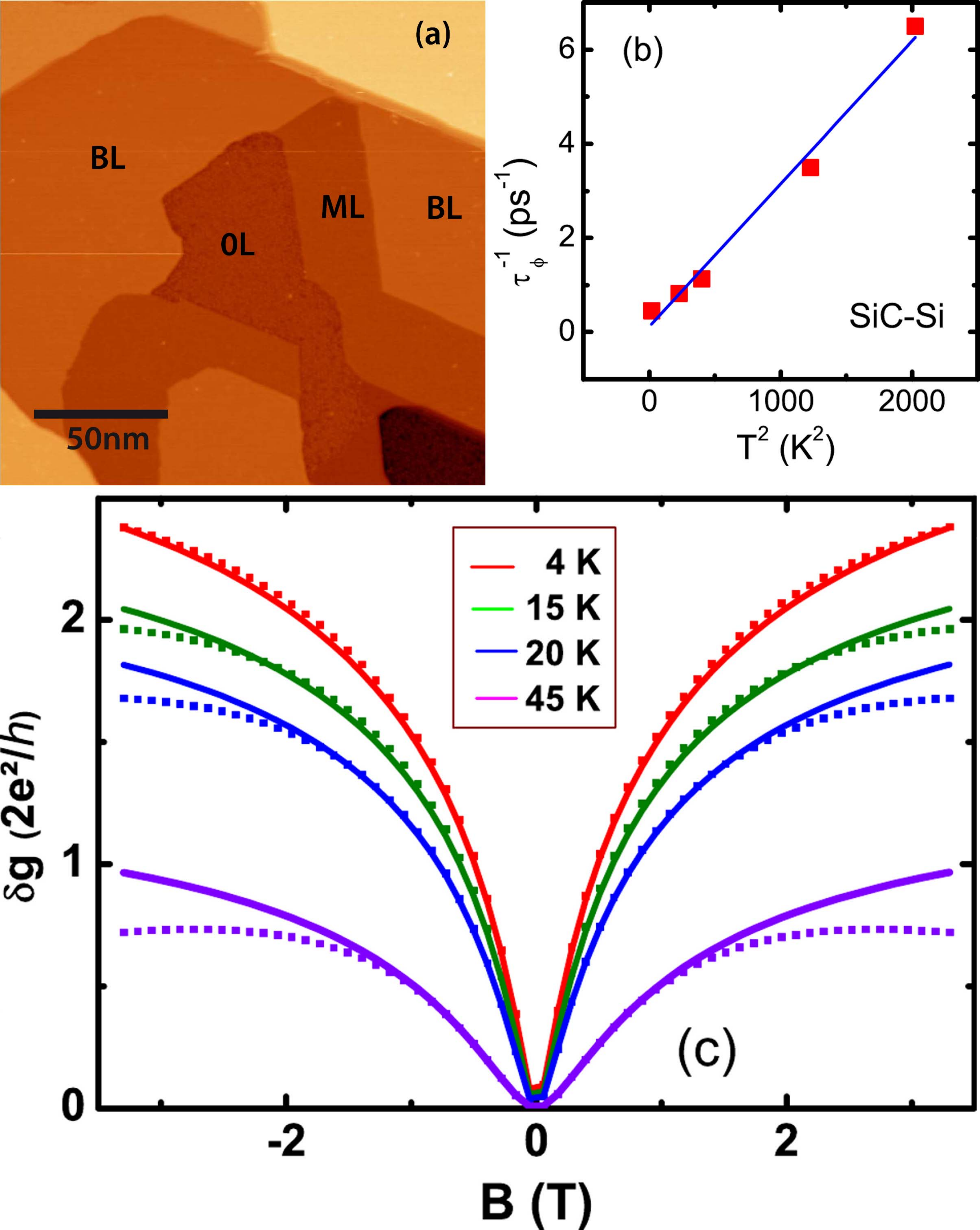}

\caption{Graphene sample grown from the SiC Si-face.
Top left (a): STM topographic image showing the terraces with
different numbers of graphene layers (the labelling as ML for
monolayer, BL for bilayer is inferred from the
step-height\cite{Lauffer08}). The average number of layers for this
sample is of the order $2$ (bilayer). Top right (b): The temperature
dependence of the phase coherence length given by the the weak
localization fits. The quadratic dependence observed suggests a
phonon-dominated dephasing process. Bottom (c): Bilayer
magnetoconductance measured at different temperatures (continuous line). The
magneto-conductance increases monotonously with field at low $T$: it
fits the expected quantum
correction to the conductance when the A-B symmetry is broken (dashed line).
} \label{UHV-grown-samples}
\end{figure}

We first consider samples grown by graphitization of the SiC
Si-face.  The growth dynamics is slow and the average number of
graphene layers can be controlled. When the number of layers exceeds
one, the layers stack as in graphite (Bernal) and break the A-B
symmetry. In the STM topographical image shown in Figure
\ref{UHV-grown-samples}-a, the domains with different number of
layers can be precisely identified and labelled. On average, this
sample has $\approx 2$ layers and a mobility of 120 cm$^2$/(V.s). 
STM studies have shown that the bilayers cover around $50 \%$ of the surface of our sample. 
The upper graphene layer is continuous over the monolayer-bilayer junction \cite{tersoff11}. 
The electron transmission between monolayer and bilayer graphene has been theoretically
studied for zigzag and armchair boundaries. Due to the presence of an evanescent wave in the bilayer graphene, 
traveling modes are well connected to each other. The transmission through the boundary is
strongly dependent on the incident angle \cite{ando10}. Nevertheless, the experimental measurements
at the atomic scale have shown a poor transmission factor of $T=1/2$ and such junctions of monolayer-bilayer
will reduce the macroscopic conductivity \cite{tersoff11}. 
In our sample ,strong  intervalley scattering was not observed at the single/bilayer edges in STM studies due to
the smooth scattering potential in the continuous surface layer \cite{mahmood12}. 
The magnetoconductance traces plotted in units of $2 e^2/h$ for
different temperatures are shown in Figure \ref{UHV-grown-samples}-c: the
continuous increase observed as a function of field saturates on the
scale of $e^2/h$. All curves can be fitted with the
weak-localization formulae \cite{McCann06,Kechedzhi07} $\delta G^{bi}=
\delta G_i-\delta G_*$
appropriate for bilayers, assuming that the phase coherence length
$L_\varphi$ is the only temperature dependent parameter. Despite the fact that the sample is inhomogeneous, we assume that 
the signal from the majority bilayer areas dominates the weak localization correction 
\footnote{The weak localization of an inhomogeneous sample of mono- and bi-layer can be written as 
$\delta G^{total}= \alpha \delta G^{mono}+\beta \delta G^{bi}$ with $\delta G^{mono}= \delta G_i+\delta G_*$ and $\delta G^{bi}= \delta
G_i-\delta G_*$. Due to the fact that the diffusion constant is smaller for the monlayer the amplitude of $\delta G^{mono}$ is smaller than the amplitude of  $\delta G^{bi}$. 
 $B^{mono}_* =\frac{\Phi_0}{4\pi D \tau^{mono}_{*}}$ which gives the magnetic field of the maximum of $\delta G^{mono}$ is larger than $B^{bi}_*$ and the effect of
$\delta G^{mono}$ is  roughly to enlarge the magnetoresistance. Nevertheless the numerical results of the fits using the bilayer formulae seems correct.}.
 The
characteristic field $B_*=0.935\:$T obtained from the fits is close
to twice $B_i=0.44\:$T. This ratio has been observed in most samples
studied. Using $B_*, B_i$, the intervalley $L_i$ and intravalley
$L_*$ diffusion lengths are found to be comparable to the size of
the flat terraces. This suggests that intervalley and intravalley
scattering are dominated by
the boundaries between domains. The time-reversed loops thus concern homogeneous domains. 
The corresponding timescales
$\tau_{i,*}=L_{\varphi,i,*}^2/(4D)$ are set by the diffusion
coefficient D for this 2D-sample. Using the fits, the dependence of
the phase breaking rate $\tau_\varphi^{-1}$ on temperature, is shown
in Figure \ref{UHV-grown-samples}-b. The $T^2$ behavior observed should be
contrasted with the linear dependence observed for the
C-face \cite{Wu07}. A quadratic increase of $\tau_\varphi^{-1}$ has
also been observed at high temperatures by Tikhonenko et
al. \cite{Tikhonenko09}.  The linear contribution due to
electron-electron dephasing appears to be quenched by the gap
induced by Bernal stacking, leaving another scattering mechanism,
probably associated with phonons, as the dominant one. For
semiconductor 2D electronic systems, the electron-phonon scattering
rate is known both from theory and experiments to increase as $T^3$.
For isolated graphene planes, different regimes \cite{Hwang08} have
been identified depending on the relative value of the temperature
compared to the Bloch-Gr\"uneisen temperature $T_{BG}=2k_F v_{\rm
ph}/k_{B}$ ($\approx 90 K$ for this sample): below $T_{BG}$, the
rate grows as $T^4$, while it is linear above \cite{Stauber07}. How
the SiC substrate affects this behavior is not known.  While the
fits (Figure \ref{UHV-grown-samples}-c) are quite accurate at low fields,
deviations can be observed at high temperature and high fields: a
negative component in the magnetoresistance traces grows at large
fields as the temperature is raised. The origin of this
classical-like behavior is not clear, but it is concomitant with the
appearance of the quadratic dephasing rate.

\subsection{\label{sec:C-face}Magneto-conductance on the C-face of 6H-SiC}

\begin{figure}[!pth]
\includegraphics[bb= 0 0 1729 2202,width=0.48\columnwidth]{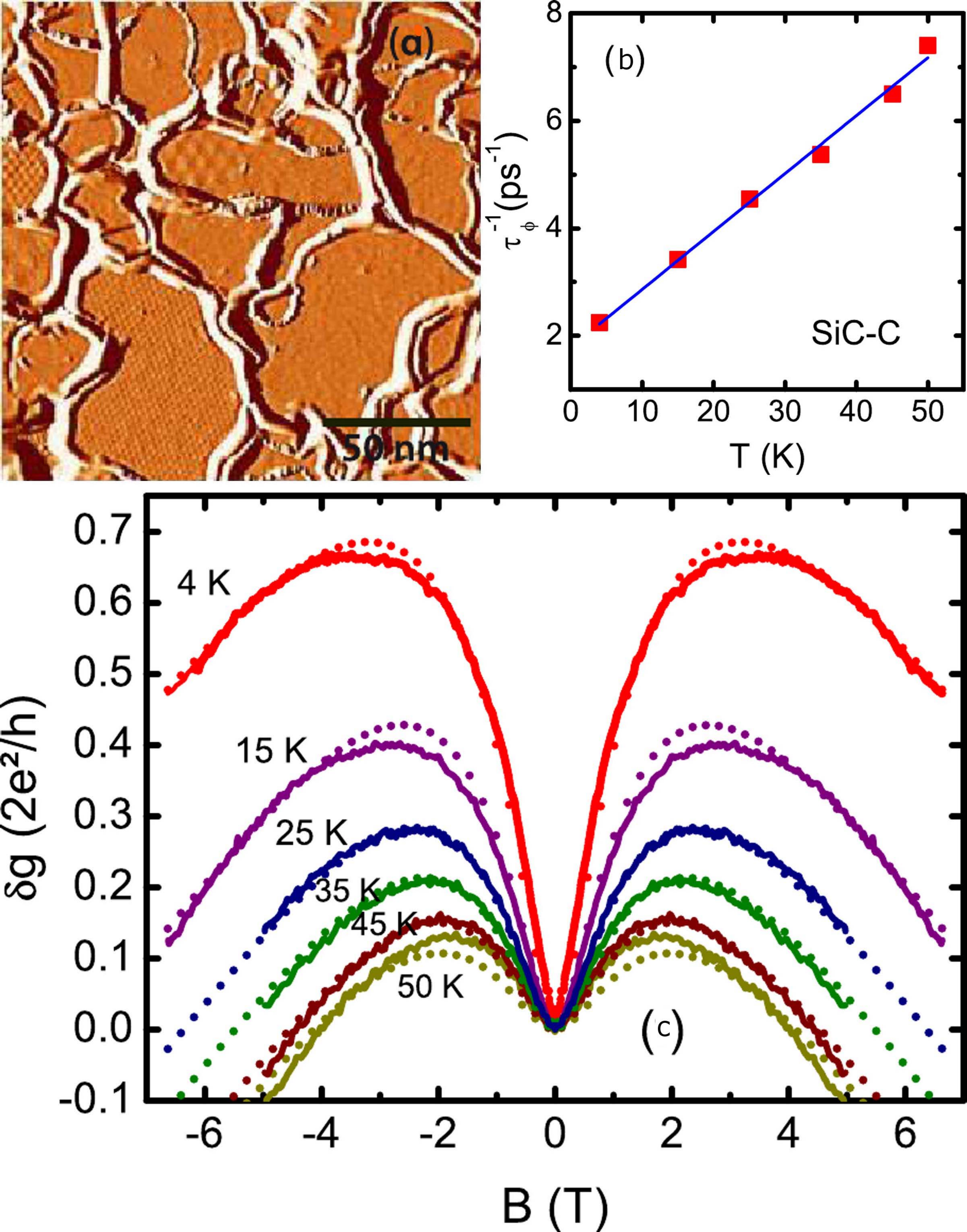}

\caption{Graphene sample grown from SiC C-face. Top left (a) :
Derivative STM image (200 nm$\times$200 nm) showing the surface morphology of a
sample graphitized in UHV. The flat terraces are of the order of 60
nm in size and are separated by ripples and folds. Top right (b):
temperature dependence of the dephasing rate on the carbon face.
Bottom (c): The magnetoconductance traces of C-face grown samples
show a weak localization dip (probing the longest coherent loops)
close to zero field and a negative magnetoconductance at higher
field (anti-localization for short loops). The crossover regime
occurs around the field $B_*\approx 2B_i$. At higher temperature,
long loops are cut-off ($L_\varphi$ decreases at higher T) reducing
the weak localization contribution at low field.  The contrast with
the Si face (Fig. \ref{UHV-grown-samples}-c) is quite clear.}
\label{UHV-grown-samplesC}
\end{figure}

Samples grown from SiC C-face have notably different morphologies
when grown in high vacuum or in an inert atmosphere. When grown in
ultra-high vacuum, flat terraces are relatively small as shown on
the topographic STM image (Figure \ref{UHV-grown-samplesC}-a top-left), typically of order 60 nm with folds and ripples at their
boundaries. When grown in an inert atmosphere, the domains (and
mobilities) are much larger (see figure \ref{C-thick}).

As long as the number of layers is small ($\le 10$),  it is possible
to analyze the magneto-conductance in term of the weak
localization-antilocalization effects discussed above, in agrement
with earlier studies \cite{Wu07, Tikhonenko09, Creeth11}.  This is illustrated
in the bottom panel of Figure \ref{UHV-grown-samplesC}-c, where the
magnetoconductance traces of a UHV grown sample, scaled in units of
$2e^2/h$, are plotted for different temperatures. All traces can be
fitted to the weak localization correction discussed in equations
\ref{intervalley} and \ref{intravalley}.  In particular, the weak
localization dip observed at low field (i.e. for the longest
coherent retrodiffusion loops) turns into a negative
magnetoconductance at higher fields (anti-localization for small
retrodiffusion loops). The characteristic field $B_*\approx 1.8\:$T
is found to be nearly twice $B_i\approx 0.72\:$T as for the Si-face
grown samples. The corresponding intervalley $L_i$ and intravalley
$L_*$ diffusion length are also of the order of the size of the flat
terraces (see the table \ref{Tab-samples}). All
traces can be fitted assuming that only $L_\varphi$ varies with
temperature. The dependence of the dephasing rate
$\tau_{\varphi}^{-1}$ with $T$ is shown in Figure
\ref{UHV-grown-samplesC}-b. The linear temperature dependence, also
observed in Wu et al. \cite{Wu07} is consistent with
Altshuler-Aronov prediction for electron-electron interactions \cite{Altshuler85,Aleiner99}
$\frac{h}{\tau_{\varphi}} = \frac{k_{B} T}{2\pi}
\frac{R_\square}{R_K} \ln \left(\frac{\pi R_K}{R_\square}\right)$
with $R_K = h/e^2$ is the quantum resitance and $R_{\square}$ is the square resistance.
The measured slope ($\approx
8\times 10^{10}$s$^{-1}$K$^{-1}$) is an order of magnitude larger
than the expected value.  Similar discrepancies are not uncommon in
other semiconducting 2DEG systems. Among the other sources of
dephasing, the Coulomb scattering of electrons in different layers
have been shown theoretically to be relevant \cite{Darancet08}.
Electron-phonon scattering may also
contribute \cite{Stauber07,Hwang08}.

\subsection{Thick graphene sample}

\begin{figure}[!h]
\includegraphics[bb= 0 0 800 600,width=0.6\columnwidth]{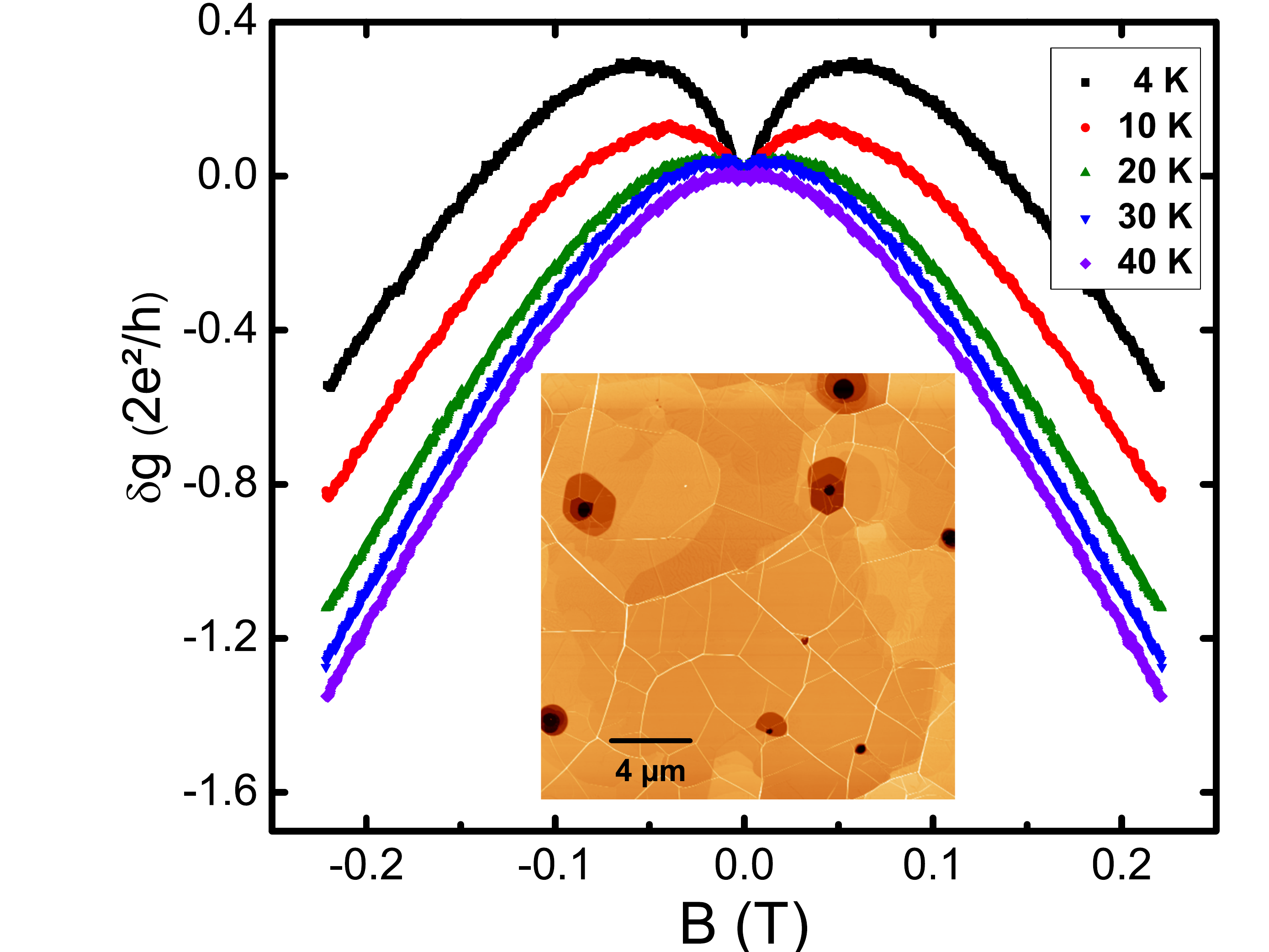}
\caption{Magneto-conductance of a thick graphene stack, annealed at
high temperature.  A narrow weak localization dip close to zero
field is observed at low temperature.  Above 0.05 T, the negative
magnetoconductance cannot be attributed to weak-localization alone.
The data is consistent with classical reduction of the longitudinal
conductance for electron with relativistic dynamics.
Inset: AFM image of the sample showing large (several microns)
domains separated by ripples.  The holes (black dots) are about
30-40 nm deep.} \label{C-thick}
\end{figure}

We now turn to samples graphitized from the SiC C-face in an inert
atmosphere which have significantly different morphologies and
transport behaviors. When graphitizing from the SiC C-face, the
number of layers increases rapidly and a larger number of graphene
layers (of order of 30-50 for the sample studied) is reached in a 10
minutes period.  After annealing at high temperature, the sample
morphology (as shown in the AFM image \ref{C-thick}-inset) shows
very large (several $\mu$m) domains separated by fold or ripples.
The magnetoconductance traces measured  at different temperatures
are shown in Figure \ref{C-thick}. A narrow weak localization dip close to
zero field is clearly seen: its width is controlled by the phase
coherence length $L_\varphi$ (of the order of 750 nm at 4 K). The
diffusion coefficient (1600 cm$^2$/s) and mobilities
($10^4$cm$^2$/(V.s)) are also found to be much larger than in the
UHV grown samples.

Within this thick sample, the conduction is no
longer limited to the layer closest to the interface with the SiC.
From the measured value of the
diffusion coefficient, we infer a large mean free path $\ell_e \approx 300
nm$ which comparable to $L_\varphi$. For the semi-classical theory, the curvature of electron trajectories usually results in a positive quadratic magneto-resistance
and so a negative magneto-conductance :
$\sigma_{xx}=\frac{\sigma_{0}}{1+(\omega_{c}\tau_L)^2}$ with $\sigma_{0}$ the conductivity without magnetic field and $\omega_{c} = v_F \sqrt{2(\frac{e}{\hbar}) B}$
 the relativistic graphene cyclotron frequency.
A fit to the magnetoconductance shown in
Fig.~\ref{C-thick} gives a scattering time $\tau_L \approx 0.05$ps
shorter than the mean-free time $\tau_e \approx 0.3$ps suggesting
other scattering processes are at play in a magnetic field (Coulomb
scattering between layers \cite{Darancet08}, tunneling between
graphene plane) which contribute to the broadening of Landau levels
and shorten the effective scattering times.
The amplitude of the negative magnetoconductance above 0.05 T shown in Fig.~\ref{C-thick}
also exceeds $e^2/h$ and cannot be attributed to quantum
interference alone. Assuming relativistic electron dynamics, the cyclotron radius
$r_c = v_F/\omega_c = \ell_B/\sqrt{2}$ becomes smaller than the mean
free path for fields exceeding ($B_c \approx 10^{-1}$ T).  As seen in
Fig.~\ref{C-thick}, the magnetoconductance turns negative above
$0.05$ T ($> B_c$).  At larger field($> 0.4T$), a linear magnetoresistance
regime is observed (the top curve in Fig. 1), which is consistent with previous
transport measurements on epitaxial graphene \cite{Friedman10}.
Several routes to linear magnetoconductance have been previously considered.
For inhomogeneous and disordered materials, a classical resistor
network model \cite{Parish03,Parish05} accounts for the linear
magnetoresistance observed in silver chalcogenides.  Another mechanism leading to a
linear magnetoconductance has been considered by Abrikosov\cite{Abrikosov99, Abrikosov00}
for layered materials and small (and zero) gap 3d-semiconductors:
in the quantum limit, when the temperature and Fermi energy
are smaller than the Landau band splitting, the magnetoresistance becomes linear.
For a pure 2d system, the requirements for quantum linear magnetoresistance coincide
with the quantum Hall regime.  On these samples, the onset
of the linear regime (0.5 T) which coincides approximatively to $\omega_c \tau_L \ge 1$
occurs well before the Shubnikov-de Haas oscillations are observed \footnote{For the sample obtained in UHV, the
poor mobility  imply that the quantum hall regime is obtained at higher magnetic fields. Thus the linear magnetoresistance regim cannot be 
reached for the range of magnetic field values used in this study}.
This indicates that tunneling between layers is larger in these thick graphene samples
compared to the few layer samples (number of layers $< 10$.  This makes the Abrikosov mechanism
the most probable explanation for the linear magnetoresitance, in agreement with Ref.\cite{Friedman10}.
From a device point of view, the transport characteristics of such thick graphene stack
are good.  Gating effects measured on such thick graphene stacks have however been found to be small.

\subsection{Discussion of the intravalley versus intervalley scattering length}

The quasiparticles of graphene can be described in the space of four-component wave functions, $|A \rangle_{K+},
|B \rangle_{K+}, |B \rangle_{K-},
|A \rangle_{K-}$ basis describing electronic amplitude on $A$ and $B$ sites and in the valleys $K+$ and $K-$.
In order to describe the microscopic scattering potentials, we introduce two sets of $4X4$ hermitian matrices : the isospin ($\vec \Sigma = ( \Sigma_{x}, \Sigma_{y}, \Sigma_{z} )$) and the pseudospin ($\vec \Lambda = (\Lambda_{x}, \Lambda{y}, \Lambda_{z} )$) \cite{McCann06,Kechedzhi07}.
 Then the electron hamiltonian in weakly disorder graphene can be parameterized as

 \begin{equation}
 V(\vec
r)=u_0(\vec r){\hat {\bf I}}+\sum_{i,j} u_{i,j}({\vec r})\Lambda_i
\Sigma_j
 (i,j\equiv x,y,z)
\label{potential}
\end{equation}

For each scattering potential
$u_{i,j}({\vec r})$, there is a microscopic scattering rate
$\tau_{ij}^{-1}$. Since $x$ and $y$ are equivalent ($\perp$), there
are only four microscopic scattering rates, $\tau_{zz}^{-1},
\tau_{z\perp}^{-1}, \tau_{\perp z}^{-1}$ and $\tau_{\perp
\perp}^{-1}$. If the sample is sufficiently disordered, it is
plausible to assume that all the potential $u_{ij}$ and scattering
rates $\tau_{ij}^{-1}$ are comparable.  In this limit, the inter and
intravalley scattering rates \cite{Kechedzhi07} $\tau_i^{-1}=
4\tau_{\perp \perp}^{-1}+ 2\tau_{z\perp}^{-1} \approx 6\tau_0^{-1}$
and $\tau_*^{-1}=\tau_i^{-1}+2\tau_z^{-1}\approx 12\tau_0^{-1}$
since $\tau_z=2\tau_{zz}^{-1}+\tau_{\perp z}^{-1}$. $B_* =\frac{\Phi_0}{4\pi D \tau_{*}}$ is found
to be twice $B_i = \frac{\Phi_0}{4\pi D \tau_{i}}$ in agreement with our experimental results, and the weak localization correction depends only
on $B_*$ as $B_\varphi \stackrel{T\rightarrow 0}{\longrightarrow}
0$: there is a universal scaling of the magnetoconductance in
$B/B_*$ and all samples-magnetoconductances collapse on this curve
at low temperatures.

\section{Conclusions}

In this paper, an overview of weak localization properties on a
variety of epitaxially grown samples has been presented.
For all types of few-layers graphene samples, the measured
characteristic lengthscales for iso and pseudo-spin diffusion
(intra/intervalley scattering) coincide with the terrace and or
domain sizes identified on the samples by STM or AFM images. For most samples,
the magnitude of the intervalley and intravalley scattering rates
have a ratio of $2$. Such a ratio is found when all scattering rates
$\tau_{ij}^{-1}$ induced by the scattering potentials (cf. Eq.~\ref{potential})
$\tau_{ij}^{-1}$ ($i,j\equiv, z,\perp$) are of similar magnitude.
 The propagation through a domain is ballistic $ l_e=V_F \tau_e$ and most of the
 diffusion occurs at the edge of the terraces where all types of scattering processes are present. 
In the case of thick samples grown in furnace, the elastic mean free path is smaller than the domains' size 
and the propagation through a domain is no more ballistic but diffusive. 
Other scattering processes are involved like coulomb scattering between layers or tunneling between graphene plane.

\begin{acknowledgements}
We acknowledge L. Magaud for her critical reading of this article.
This work is supported by the French National Research Agency (GraphSiC and XPgraphene projects) and the Nanosciences Foundation (DISPOGRAPH project).

\end{acknowledgements}
            
\bibliographystyle{apsrev4-1}
\end{document}